\documentclass[8pt,aps]{revtex4}
\usepackage{amssymb}
\usepackage{latexsym}

\usepackage{epsfig}

\usepackage{dcolumn}
\usepackage{amsmath}
\usepackage{amsfonts}
\usepackage{bm}

\newcommand{\be}{\begin{equation}}
\newcommand{\ee}{\end{equation}}
\newcommand{\bea}{\begin{eqnarray}}
\newcommand{\eea}{\end{eqnarray}}

\newcommand{\p}{\partial}
\newcommand{\s}{\sigma}

\newcommand{\la}{\langle}
\newcommand{\ra}{\rangle}
\newcommand{\rd}{\mbox{d}}
\newcommand{\ri}{\mbox{i}}
\newcommand{\re}{\mbox{e}}

\begin{document}
\title{Topological  Kondo effect in star junctions of Ising magnetic chains. Exact solution. }
\author{A. M. Tsvelik}
\affiliation{ Department of Condensed Matter Physics and Materials Science, Brookhaven National Laboratory, Upton, NY 11973-5000, USA}
 \date{\today } 
\begin{abstract} 
In this paper I present a conjecture for the Bethe ansatz equations for  the  model describing  a star junction of $M$ quantum critical Ising chains. For $M>3$ such model exhibits the so-called topological Kondo effect [B. Beri and N. R. Cooper, Phys. Rev. Lett. {\bf 109}, 156803 (2012)] related  to existence of   Majorana zero energy modes  at the junction. These modes are of a topological nature; they non-locally encode an SO(M) "spin" which is screened by the collective excitations of the chains. For certain values of $M$ the model is equivalent to the Kondo models with known exact solution. These cases are used to check the validity of the conjecture. It is  demonstrated that the model behaves differently for $M$ even and odd; in the former case the model has a Fermi liquid and the latter case corresponds to a non-Fermi liquid infrared fixed point.  
\end{abstract}


\maketitle
\section{Introduction}
 
 In this paper  I suggest  a Bethe ansatz solution  of the  model describing a star junction of $M$ quantum critical Ising chains. For $M\geq 3$ this model describes the so-called topological Kondo effect (the concept was introduced in \cite{cooper}) related  to existence of   Majorana zero energy modes (MZEMs) at the junction. As was first established in \cite{goshal}, such modes emerge as a necessary ingredient of the  fermionic description of the quantum Ising model with a boundary. Later in \cite{crampe},\cite{Y} where the Jordan-Wigner transformation for star graphs was introduced, it was shown that MZEMs are also present in  the lattice description. The  operators describing these zero modes compose a spinor representation of the SO$(M)$ group and thus can be thought of as components of a kind of SO$(M)$ ``spin''. Gapless collective excitations of the critical Ising chains are scattered by this spin giving rise to a new type of Kondo effect.  It is remarkable that emerging as a book-keeping device, the MZEMs are thus elevated to the status of dynamical degrees of freedom.

 In my previous  paper \cite{Y} it was established that  a junction  of three quantum critical Ising models coupled together at one point by ferromagnetic interactions is equivalent to the spin sector of the two-channel Kondo model where the impurity spin has magnitude S=1/2.  This model is known to possess  a non-Fermi liquid fixed point \cite{blandin} with a residual ground state entropy \cite{wiegmann},\cite{andrei},\cite{tsvelik} and  power law singularities in the thermodynamic quantities. Below, I make a conjecture about  the exact  solution for arbitrary $M$ and demonstrate that there is a qualitative difference between $M$ odd and even. In the former case the low temperature behavior is quantum critical and in the latter case it is not. This is the central result of the paper.

 In Section II I present a derivation of the topological Kondo model for a star junction of critical Ising chains. Its content mostly coincides with the one of \cite{Y}; it is  included to make the paper self-contained. In Section III I present the Bethe ansatz equations for $M>3$ and discuss the properties of their solutions. In Section IV the results extracted from the Bethe ansatz are compared with the ones extracted from Conformal Field Theory. The paper has  Conclusions and Acknowledgements sections and  two  appendices where I discuss Bethe ansatz thermodynamics and the results for the $M=3$ case.

\section{From spins to fermions. Jordan-Wigner transformation}

The Hamiltonian  describes $M$ species of  quantum Ising chains 
coupled together in a junction:
\bea
&& H = -\sum_{p>q} J_{pq}\s^x_p(1)\s^x_q(1)  + \sum_{p=1}^M H_{Ising,p}, \label{Y}\\
&& H_{Ising,p} = \sum_{j=1}^N\Big[-J\s^x_p(j)\s^x_p(j+1)  + h\s^z_p(j)\Big]. \label{Ising}
\eea
It is assumed that $0< J_{pq} << J$.  Models (\ref{Ising}) are equivalent to noninteracting Majorana fermions by means of Jordan-Wigner transformation. To extend this formalism for Ising models on star graphs one should follow \cite{crampe},\cite{Y} and  introduce  Klein factors  to mark  the different chains:
 \bea
&& c_p(j) = \gamma_p\left(\prod_{k=1}^{j-1}\s^z_p(k)\right)\s^-_p(j), \label{ferm}\\
&& \s^-_p(j) = \gamma_p c_p(j)\exp\left[\ri\pi\sum_{k=1}^{j-1}c^+_p(k)c_p(k)\right], ~~\s^z_p(j) = c^+_p(j)c_p(j)- 1/2, \nonumber
\eea
where  Klein factors $\gamma_p$ satisfy  Clifford algebra 
\be
\{\gamma_i,\gamma_j\} = 2\delta_{ij}, ~~i,j=1,...,N, \label{Cliff}
\ee
 and commute with all  spin operators. Anticommutativity of $\gamma_i$'s  establishes  commutativity of the spin operators located at  different chains. Hence, the emergence of  MZEMs in the fermionic formulation of the problem is quite natural.  Notice also that the boundary spins  have a particularly simple expression in terms of fermions:
\be
\s^x_p(1) = \gamma_p[c_p(1)- c^+_p(1)],  \label{boundary}
\ee
This equation is valid for a single semi-infinite chain and thus  is a lattice generalization of Eq.(4.25) of \cite{goshal}. 

Substituting (\ref{ferm}) into (\ref{Ising},\ref{Y}) we obtain the following fermionic Hamiltonian:
\bea
&& H = \sum_{p=1}^3 H_{Ising,p} +V\label{Hferm}\\
&& H_{Ising} = \sum_{j=1}\Big\{-J[c^+(j+1)-c(j+1)][c^+(j)+c(j)] +  hc^+(j)c(j)\Big\}\label{Isferm}\\
&& V =\sum_{p>q} J_{pq}\gamma_p\gamma_q[c_p(1)-c^+_p(1)][c_q(1)-c^+_q(1)] \label{Vferm}
\eea
As we see, the MZEMs represented by the Klein factors do not appear in the expressions for the bulk Hamiltonians (\ref{Isferm}). They appear only in the term describing the junction and for this reason their combination can be interpreted as a quantum degree of freedom located at the junction. 

 To check the formalism it is instructive to consider first the trivial case of two chains which can be solved by the conventional Jordan-Wigner transformation. The interaction in that case is just
\bea
J_{12}\gamma_1\gamma_2[c_1(1)-c^+_1(1)][c_2(1)-c^+_2(1)].
\eea
Since $\ri\gamma_1\gamma_2 = \pm 1$ the Klein factors can be dropped. A possible minus sign can be removed  by the particle-hole transformation on one of the chains. Thus we see that the problem of two chains is equivalent to the one of a single chain with a weak link. The Klein factors are redundant as expected. For a junction of $M \geq 3$ chains this is no longer the case. As we will see, the Klein factors then play a prominent role. This is a subject of the subsequent sections.

\section{The continuum limit for $M \geq 3$}

 To establish the aforementioned relation between the  model describing a junction of critical Ising chains and the Kondo model one needs to switch to the continuous description of model (\ref{Hferm},\ref{Isferm}) valid for $|J-h| << J$. It is convenient to introduce the following bulk  Majorana fermions 
\bea
\xi(j )= c^+(j) + c(j), ~~ \rho(j) = \ri[c^+(j)-c(j)],
\eea
 then pass to the continuum limit and introduce the right- and the left-moving modes $\chi_{R,L}(x) = [\rho(x=ja_0) \pm\xi(x=ja_0)]/\sqrt {2a_0}$, where $a_0$ is the lattice constant. The resulting continuum limit Lagrangian for a single chain with a boundary (\ref{Isferm}) is then 
\bea
 && L = \int_{0}^{\infty}\rd x \Big[\frac{1}{2}\chi_R(\p_{\tau} - \ri v\p_x)\chi_R + \frac{1}{2}\chi_L(\p_{\tau} + \ri v\p_x)\chi_L +  \ri m\chi_L\chi_R\Big] + \frac{\ri}{2}\gamma\p_{\tau}\gamma,\label{cont}
\eea
where $v = Ja_0, ~~ m = J-h$. It is augmented by the  free boundary condition: 
\bea
\chi_R(0) = \chi_L(0) \label{free}
\eea
This condition corresponds to the fact that the boundary spin (\ref{boundary}) is not fixed. Lagrangian (\ref{cont}) coincides with the one obtained for the Ising model with the boundary in \cite{goshal}. 

 Let us now set $m=0$. With boundary condition (\ref{free}) one can introduce a chiral fermion field $\chi(x) = \theta(x)\chi_R(x) +\theta(-x) \chi_L(-x)$ and extend the integration in (\ref{cont}) to the entire $x$-axis.  This can be done safely since $\chi(x)$-field is continuous at $x=0$. Taking into account  interaction term (\ref{Vferm})  and switching back to the Hamiltonian formalism, we arrive at the following effective theory for the star junction of $M$ critical Ising chains:
\bea
 H_{eff} = - \sum_{p<q} J_{pq}\gamma_p\gamma_q \chi_p(0)\chi_q(0) + \frac{\ri}{2}\sum_{p=1}^{M}\int_{-\infty}^{\infty} \rd x v_p\chi_p\p_x\chi_p , \label{KondoT}
\eea
where velocities $v_p$ may be different for different chains. The fermionic bilinears in (\ref{KondoT}) are  the O$_1$(M) Kac-Moody currents. These currents represent a particular case of the O$_k$(M) currents:
\[
{\cal J}^{pq} = \sum_{j=1}^k\chi_p^{(j)}\chi^{(j)}_q.
\]
The additional index $j$ has no meaning for the Ising chains, but model (\ref{Kondo}) allows such generalization and I find it instructive to discuss the case of  arbitrary $k$. The case $k=2$ was discussed in \cite{cooper} and the star junction of spin S=1/2 XX models discussed in \cite{crampe}.

\section{Bethe ansatz}

\subsection{M=3 and M=6}

As was established in \cite{Y} (see also Appendix A), $M=3$ case is equivalent to the spin subsector of the 2-channel S=1/2 Kondo problem and as such is quantum  critical. For $M>3$  the problem is likely to be integrable when all dimensionless couplings $J_{pq}/\sqrt{v_pv_q}$ are equal:
\bea
H_{eff} = \frac{\ri v}{2}\int_{-\infty}^{\infty} \rd x \chi_p\p_x\chi_p + G \gamma_p\chi_p(0)\gamma_q\chi_q(0). \label{eff2}
\eea
For that case the Bethe ansatz equations have not been derived, but I suggest a plausible conjecture which can be checked against known facts. I consider $M=6$ when SO(6)$\sim$ SU(4)  as control cases. Here we deal with the SU(4) Kondo effect for which the Bethe ansatz equations are known for a certain class of the group representations \cite{su4},\cite{reshetikhin}. 

Model (\ref{eff2}) remains integrable in the presence of "magnetic" fields attached to bilinears of local Majoranas:
\bea
\delta H = \ri\sum_{(ij)} t_{ij}\gamma_{i}\gamma_j, \label{Zeeman}
\eea
where the sum includes non-overlapping pairs of chains. Although in the original spin system these fields do not correspond to any local operators, they constitute an important mathematical device in the analysis of the problem. Indeed, perturbation (\ref{Zeeman}) is  relevant and by applying a field to a given pair of chains one initializes a continuous  RG flow from the $M$-chain to the $(M-2)$-chain problem.  This fact already enables one to establish a difference between even and odd $M$. Indeed, by application of suitable fields one can arrange a flow from any even $M$ to $M=2$ which is, as was demonstrated in Section II, corresponds to a single chain with a weak link and hence is  a trivial Fermi liquid fixed point. On the other hand, odd $M$'s flow to $M=3$ which is the 2-channel Kondo model  QCP \cite{Y} and further to $M=1$ which is a non-interacting system with the zero energy Majorana mode playing role of an auxiliary field. The analytical study of the flows can be used  to check the conjectured Bethe ansatz equations. Namely, for even $M$ the application of the suitable Zeeman field must generate the low energy theory of $M=4$ which exact solution is known. Likewise in the  odd $M$ case the Bethe ansatz equation must allow its reduction to the $M=3$ case analysed in \cite{Y}.

 Consider $M=6$. The fermionic current operator transforms according to  the vector representation of the SO(6) group. The level of the Kac-Moody algebra is $k=1$. This representation is isomorphic to the representation of the SU(4) group with the Young tableaux with one column of two boxes.  The impurity operator transforms according to  the spinor representation of SO(6) which is isomorphic to the representation of the SU(4) which Young tableaux consisting of one box. Now following \cite{reshetikhin} one can use 
these facts to write down the Bethe ansatz equations: 

\bea
&& \prod_{b=1}^{M_2}e_{1}(\lambda_a^{(1)} -\lambda_b^{(2)})e_1(\lambda_a^{(1)} - 1/g) = \prod_{b=1}^{M_1}e_2(\lambda_a^{(1)} -\lambda_b^{(1)}),\\
&& [e_1(\lambda_a^{(1)})]^N\prod_{b=1}^{M_2}e_{1}(\lambda_a^{(2)} -\lambda_b^{(1)})\prod_{b=1}^{M_3}e_{1}(\lambda_a^{(2)} -\lambda_b^{(3)}) = \prod_{b=1}^{M_2}e_2(\lambda_a^{(2)} -\lambda_b^{(2)})\nonumber\\
&& \prod_{b=1}^{M_2}e_{1}(\lambda_a^{(3)} -\lambda_b^{(2)}) = \prod_{b=1}^{M_3}e_2(\lambda_a^{(3)} -\lambda_b^{(3)})\nonumber\\
&& E = \sum_{a=1}^{M_2} \frac{1}{2\ri}\ln e_1(\lambda_a^{(2)})\nonumber, 
\eea
where $g = G/2\pi v$ and 
\[
e_n(x) = \frac{x-\ri n/2}{x+ \ri n/2}.
\]
Now one can follow the standard procedure and to derive from these discrete  equations  Thermodynamic Bethe ansatz (TBA) equations (see Appendix B).  In the scaling limit $T_K/J \rightarrow 0$, TBAs are the same as for the SU(4) Kondo model with the bulk transforming according to  the single box representation \cite{su4}: 
\bea
&& F_{imp} = - T\sum_{j=1}^3\int f_j[x + \frac{1}{\pi}\ln(T_K/T)]\ln(1 + \re^{\phi_1^{(j)}(x)}), \\
&& \ln(1 + \re^{-\phi_n^{(j)}}) - {\cal A}_{jk}*C_{nm}*\ln(1 + \re^{\phi_m^{(k)}})=  \delta_{n,1}\sin(\pi j/4)\re^{-\pi x/2} , ~~j,k =1,2,3;\nonumber\\
&& ~~ n,m =1, 2,...\\
&& \lim_{j\rightarrow \infty} \frac{\phi_n^{(j)}}{n} = H_j/T.
\eea
where $H_j$ are "magnetic" fields and 
\[
f*g(x) = \int_{-\infty}^{\infty}f(x-y)g(y)\rd y
\]
and the Fourier transforms of the kernels are 
\bea
&& s(\omega) = (2\cosh\omega/2)^{-1}, ~~ f_j(\omega) = {\cal A}_{j,1}*s = \frac{\sinh[(2-j/2)\omega]}{\sinh 2\omega}, \nonumber\\
&& C_{nm} = \delta_{nm} - s(\delta_{n.m-1} + \delta_{n,m+1}), \nonumber\\
&& {\cal A}_{jk} = 2\coth(\omega/2)\frac{\sinh[(2 - \max{j,k}/2)\omega]\sinh[\min(j,k)\omega/2]}{\sinh2\omega}
\eea
The Kondo temperature $T_K \sim J\exp(- 1/4g)$.

 These equations were analyzed in \cite{su4}; they describe Fermi-liquid-like thermodynamics as is expected for even $M$. The ground state is a simple singlet, there is no residual degrees of freedom as for the $M=3$ case.  The difference in the low energy behavior for $M=3$ (non-FL) and $M=6$ (FL) is consistent with the above statement concerning the difference between even and odd $M$.

\subsection{Bethe ansatz for arbitrary M}

 A general form of the Bethe ansatz  equations is dictated by the symmetry considerations and can be found  in \cite{reshetikhin}. To fix such details as a position of the driving terms one needs an additional information which can be obtained by considering limiting cases where the solution is already known. 

 Here I will provide the Bethe ansatz equations for general $k$. For the O$_k$(2N+1) model we suggest the following:
\bea
&& [e_k(\lambda_a^{(1)})]^L\prod_{b=1}^{M_2}e_1(\lambda_a^{(1)} - \lambda_b^{(2)}) = \prod_{b=1}^{M_1}e_2(\lambda_a^{(1)} - \lambda_b^{(1)});\nonumber\\
&& ...\nonumber\\
&& \prod_{b=1}^{M_{N-1}}e_1(\lambda_a^{(N)} - \lambda_b^{(N-1)})e_{1/2}(\lambda_a^{(N)} -1/g) = \prod_{b=1}^{M_N}e_1(\lambda_a^{(N)} - \lambda_b^{(N)}) \label{BA5}
\eea
 The impurity is in the spinor representation and hence the corresponding driving term is in the last equation. These equations successfully incorporate the limiting case $M=3$ considered in Appendix B. 

For the O$_k$(2N +2) model equations we have 
\bea
 && [e_k(\lambda_a^{(1)})]^L\prod_{b=1}^{M_2}e_1(\lambda_a^{(1)} - \lambda_b^{(2)}) = \prod_{b=1}^{M_1}e_2(\lambda_a^{(1)} - \lambda_b^{(1)});\nonumber\\
&& ...\nonumber\\
&& \prod_{b=1}^{M_{N-2}}e_1(\lambda_a^{(N-1)} - \lambda_b^{(N-2)})\prod_{b=1}^{M_{N-1}}e_1(\lambda_a^{(N-1)} - \lambda_b^{(N)})\prod_{b=1}^{M_{N+1}}e_1(\lambda_a^{(N-1)} - \lambda_b^{(N+1)}) = \prod_{c=1}e_2(\lambda_a^{(N-1)} - \lambda_b^{(N-1)}) \nonumber\\
&& \prod_{b=1}^{M_{N-1}}e_1(\lambda_a^{(N)} - \lambda_b^{(N-1)})e_{1}(\lambda_a^{(N)} -1/g) = \prod_{b=1}^{M_N}e_2(\lambda_a^{(N)} - \lambda_b^{(N)}) \nonumber\\
 && \prod_{b=1}^{M_{N-1}}e_1(\lambda_a^{(N+1)} - \lambda_b^{(N-1)}) = \prod_{b=1}^{M_{N+1}}e_2(\lambda_a^{(N+1)} - \lambda_b^{(N+1)}) \label{BA6}
\eea
For $N=2$ ($M=6$) these equations coincide with the ones for the SU(4) model analyzed in the previous Subsection. 

We have to check the suggested equations against various known facts, such as $M \rightarrow M-2$ flows. For simplicity we confine our discussion to the O(5) group ($N=2$). Then  equations for the root densities are
\bea
&& s*A_{n,k} = \tilde\s_n + A_{nm}*\s_m - [s*A_{2n,m}(\omega/2)]*\rho_m\label{roots}\\
&&  \frac{1}{L}[a_n(\omega/2)]\re^{\ri\omega/g}= -[s(\omega/2)*A_{n,2m}(\omega/2)]*\s_m + \tilde\rho_n + [A_{nm}(\omega/2)]*\rho_m, \label{rootrho} 
\eea
where the convolution sign means 
\[
[f(\omega)]*g = \frac{1}{2\pi}\int \rd y g(y)\Big[\int\rd\omega \re^{-\ri\omega(x-y)}f(\omega)\Big]
\]
and 
\[
A_{nm}(\omega) = \coth\omega/2\Big[\re^{-|n-m||\omega|/2} - \re^{-(n+m)|\omega|/2}\Big].
\]
The vacuum consists of $\s_k$ and $\rho_{2k}$. All other densities (including the densities of holes) are zero. 

The first important observation is the flow $M=5 \rightarrow M=3$. Application of the "magnetic" field creates holes in the $k$-th string ($\tilde\s_k \neq 0$ at $x>Q$, where $Q$ is determined by the field) and at the same time pushes all other strings in $\s$ to finite energies. Therefore at sufficiently low energies one can replace the term in (\ref{rootrho}):
\bea
[s(\omega/2)*A_{n,2m}(\omega/2)]*\s_m  \rightarrow [s(\omega/2)*A_{n,2k}(\omega/2)]*\s_k^{(0)} ,
\eea
where $\s_k^{(0)}$ is the vacuum value of the corresponding density. Then inverting the kernel in (\ref{rootrho}) one obtains 
\bea
\frac{1}{L}\delta_{n,1}s(x-1/g) + \delta_{n,2k}s*\s_k^{(0)}(x/2) = C_{nm}*\tilde\rho_m(x) + \rho_n(x)
\eea
At large $x$ he bulk driving term in the left hand side is approximated as 
\bea
s*\s_k^{(0)}(x) \approx A\re^{-\pi x}, ~~ A =\int_{-\infty}^Q\re^{\pi y}\s_k^{(0)}(y/2)\rd y.
\eea
and we obtain the scaling limit of the Bethe ansatz equations for the densities of the SU$_{2k}$(2) Kоndo model with the impurity of spin S=1/2:
\bea
\frac{1}{L}\delta_{n,1}s(x-1/g) + \delta_{n,2k}A\re^{-\pi x} = C_{nm}*\tilde\rho_m(x) + \rho_n(x).
\eea
The correct value of the spin is important, it justifies our choice of the impurity representation in (\ref{BA5}).

 The TBA equations are given below.  
There are two types of energies: $\phi_n$ related to $\rho_n$ and $\xi_n$ related to $\s_n$.
\bea
&& \phi_{2n+1} = s_{1/2}*\ln(1+\re^{\phi_{2n}})(1+\re^{\phi_{2n+2}}), ~~n=0,1,2,...\nonumber\\
&& \phi_{2n} = - \delta_{n,k}\re^{-2\pi x/3} + s_{1/2}*\ln(1+\re^{\phi_{2n-1}})(1+\re^{\phi_{2n+1}}) -  \frac{s_{1/2}}{1-s}*\Big[s_{1/2}C_{2n,m}*\ln(1+\re^{\phi_m}) + C_{2n,m}*\ln(1+\re^{\xi_m})\Big],\nonumber\\
&&-\ln(1+\re^{-\xi_n}) = -\frac{C_{nm}}{1-s}*\ln(1+\re^{\xi_{m}}) - \frac{s_{1/2}*C_{nm}}{1-s}*\ln(1+ \re^{\phi_{2m}}) - \delta_{n,k}\re^{-2\pi x/3}.\label{F5}
\eea
\bea
 && F_{imp} = - T\int \rd x \Big\{s_{3/2}[x + \frac{3}{2\pi}\ln(T_K/T)]\ln[1 + \re^{\phi_2(x)}] +\nonumber\\
&& s_{1/2}[x + \frac{3}{2\pi}\ln(T_K/T)]\ln[1 + \re^{\phi_1(x)}] +\nonumber\\
&&  s_{1/2}*s_{3/2}[x + \frac{3}{2\pi}\ln(T_K/T)]\ln[1+ \re^{\xi_1(x)}]\Big\}. \label{F5a}
\eea 
where $s_{n/2} = s(n\omega/2)$. 

 In the limit $T \rightarrow 0$ the leading contribution to the impurity free energy comes from $x \rightarrow -\infty$. Since in this limit both $\phi_2$ and $\xi_1$ are equal to $-\infty$, the ground state entropy is determined by $\phi_1(-\infty) =0$. From (\ref{F5a}) we get 
\be
S_{imp}(0) = \ln\sqrt 2.
\ee
 
 Below we consider the ground state "magnetization" $S_{ab} = \ri\la\gamma_a\gamma_b\ra$.
At $T=0$ the equations for the nonvanishing densities $\s^{(\pm)} \equiv \tilde\s_k,\s_k$ and $\rho^{(\pm)} \equiv \tilde\rho_{2k},\rho_{2k}$ are 
\bea
&& \s^{(-)} + \Big[\frac{\cosh(\omega/2)\sinh\omega \re^{k|\omega|}}{\cosh(3\omega/2)\sinh(k\omega)}\Big]\s^{(+)} + \Big[\frac{\sinh\omega \cosh(\omega/2)\re^{k|\omega|}}{2\sinh(k\omega)\cosh(3\omega/2)}\Big]\rho^{(+)} = \nonumber\\
&& \frac{\cosh(\omega/2)}{\cosh(3\omega/2)} + \frac{\re^{\ri\omega/g}}{L}\frac{\sinh\omega}{2\cosh(3\omega/2)\sinh(k\omega)}.\label{first1}\\
&& \rho^{(-)} + \Big[\frac{\sinh\omega \cosh(\omega/2)\re^{k|\omega|}}{2\sinh(k\omega)\cosh(3\omega/2)}\Big]\s^{(+)} + \Big[\frac{\sinh(\omega/2)\cosh\omega \re^{k|\omega|/2}}{2\sinh(k\omega/2)\cosh(3\omega/2)}\Big]\rho^{(+)} = \nonumber\\
&& \frac{1}{2\cosh(3\omega/2)} + \frac{\re^{\ri\omega/g}}{L}\frac{\sinh(\omega/2)\cosh\omega}{\sinh(k\omega)\cosh(3\omega/2)}\label{last2}.
\eea

To calculate a typical response  to the applied field for k=1 it is sufficient to consider a situation when the holes appear just in one density, for instance in $\rho$. Then I use the 2nd equation in (\ref{last2}):
\bea
\rho^{(-)}(\omega) + \Big[\frac{\cosh\omega \re^{|\omega|/2}}{2\cosh(3\omega/2)}\Big]\rho^{(+)}(\omega) =  \frac{\re^{-\ri B\omega}}{2\cosh(3\omega/2)} + \frac{\re^{\ri\omega(1/g-B)}}{L}\frac{\cosh\omega}{2\cosh(\omega/2)\cosh(3\omega/2)}\label{susc},
\eea
where $\tilde\rho$ is nonzero on $(B,\infty)$ and the integration limit $B$ is determined by the condition that the bulk part of the magnetization is equal to $H/2\pi$.

Now $\rho^{(-)}$ and $\rho^{(+)}$ are analytical in lower and upper half planes of $\omega$ respectively and (\ref{susc}) is  the  Wiener-Hopf equation. Solving it  together with the condition for $B$ I get  
\bea
S_{ab} \sim \frac{\ri}{2\pi}\int \frac{\rd\omega}{\omega - \ri 0}\re^{-\frac{3i\omega}{\pi}\ln(T_H/H)}\frac{\cosh\omega}{G^{(-)}(\omega)\cosh(3\omega/2)\cosh(\omega/2)}, 
\eea
where $T_H/T_K = 4/G^{(-)}(-\ri\pi/3)$ and 
\[
G^{(-)}(\omega) = \frac{\Gamma(1/2+3\ri\omega/2\pi)}{\sqrt 2\Gamma(1/2 + \ri\omega/\pi)}\Big(\frac{3(\ri\omega +0)}{2\pi}\Big)^{-3\ri\omega/2\pi}\Big(\frac{(\ri\omega +0)}{\pi}\Big)^{\ri\omega/\pi}\re^{\ri\omega/2\pi},
\]
is analytic in the lower half plane. 
At small fields one can bend the integration contour in the lower half plane to obtain 
\bea
S_{ab} \sim \frac{2}{\sqrt 3 \pi G^{(-)}(-\ri\pi/3)}\frac{H}{T_H} - \frac{1}{\pi^2G^{(-)}(-\ri\pi)}\Big(\frac{H}{T_H}\Big)^3\ln(T_H/H) +...
\eea
In fact, the non-Fermi liquid contribution to the specific heat or magnetic susceptibility becomes dominant only  at $k\geq 3$ (the logarithmic susceptibility). The Fermi liquid behavior at $k=1$  goes well with our understanding of the Kondo effect.

\section{Bethe ansatz and Conformal Field Theory}

 Although Bethe ansatz equations can be used to extract more thermodynamic properties, I will not present these calculations here. Instead I use the fact that for odd $M$ the ground state is quantum critical and employ the conformal field theory. 

 The quantum critical point is described by the O$_1$(2N+1) Wess-Zumino-Novikov-Witten (WZNW) boundary conformal field theory perturbed by the irrelevant operator in the form of the first Kac-Moody descendant of the primary field in the adjoint representation \cite{knizhnik},\cite{AL}:
\bea
L_{eff} = W[SO_1(2N+1)] + \tilde g J^{\alpha}_{-1}\Phi^{\alpha}(0), \label{eff}
\eea
where $W$ is the WZNW Lagrangian of the bulk,  $\Phi^{\alpha}$ ($\alpha = 1,...N(2N+1)$) is the chiral component of the primary field in the adjoint representation and $\tilde g \sim T_K^{-1 +1/2N}$ is a coupling constant. The chiral operator $\Phi$ is  multiplied by  certain Klein factors to make its correlation functions single valued. The operator $J_1\Phi$ is generated by the fusion of the current with the primary field in the adjoint representation. The scaling dimension of the latter field is given by \cite{knizhnik}
\be
h_{ad} = \frac{c_v}{k+c_v} = \frac{2N-1}{2N},
\ee
($c_v = 2N-1$ is the Coexter number for the O(2N+1) group) and the scaling dimension of its Kac-Moody descendant is $h_{irr} = 1 +h_{adj} = 2- 1/2N$
which leads to a singular correction to the  impurity specific heat 
\be
C_{imp} \sim (T/T_K)^{2-1/N}, ~~ N >1, 
\ee
and $(T/T_K)\ln(T_K/T)$ for $N=1$. For $N>1$ this a {\it subleading} correction to the FL specific heat $\sim T/T_K$. The corresponding non-FL contribution to the zero-T "magnetization" is
\be
\delta S \sim H^{2N-1}\ln(T_H/H),
\ee
that is appears on in high orders of the $H/T_H$ expansion.

 For the O$_k$(2N+1) WZNW model the central charge of the part of the bulk interacting with the impurity is 
\be
C = \frac{kN(2N+1)}{k + 2N-1}.
\ee
For $k=1$ it is $N +1/2$ (the number of the Majorana fermion species divided by 2), which means that the entire bulk takes part in the interaction with the impurity.  This makes it possible to rewrite the effective Lagrangian  (\ref{eff}) in a slightly more explicit form:
\bea
L_{eff} = \frac{\ri}{2}\int_{-\infty}^{\infty}\sum_{a=1}^{2N+1}\rd x(\chi_a\p_{\tau}-\ri\p_x)\chi_a + g\sum_{a>b}\chi_a(0)\chi_b(0)\Phi^{[a,b]}(0),
\eea
where $[a,b]$  is an  antisymmetric combination and operator $\Phi$ is nonlocal with respect to the fermions. 


 \section{Conclusions and Acknowledgements}

 The Bethe ansatz equations for the model of a star junction of $M$ critical quantum Ising  chainssuggested in this paper have been checked against various requirements such as $M \rightarrow M-2$ flow. At low temperatures $T << T_K$ the suggested solution yields the same results for the theormodynamics as  conformal field theory. It was demonstrated that the low temperature behavior of a star junction of $M$ critical quantum Ising model chains crucially depends on whether $M$ is even or odd. In the latter case we have non-Fermi liquid quantum critical point, though non-analytic corrections to the thermodynamic quantities become subleading to the analytic ones for $M>3$.

 I am  grateful to B. L. Altshuler, L.B. Ioffe, A. A. Nersesyan and V. Kravtsov for long discussions and  interest to the work. The work  was  supported  by US DOE under contract number DE-AC02 -98 CH 10886.

\begin{appendix}
\section{ $M=3$ junction}

For $M=3$ one can introduce a fixtitious spin S=1/2 operator using the identity \cite{martin}:
\be
S^p = \frac{\ri}{4}\epsilon_{pqt}\gamma_q\gamma_t, \label{BS}
\ee
It was shown in \cite{Y} that for nearly critical chains $|J-h| << J$ the MZEMs  are subject of intense screening by  gapless bulk modes and  model (\ref{Isferm},\ref{Hferm}) is equivalent to the overscreened two-channel Kondo model.  The two-channel Kondo model has been extensively studied and a plethora of nonperturbative results have been obtained for it by means of Bethe ansatz \cite{andrei},\cite{wiegmann},\cite{tsvelik} and conformal field theory (see, for instance, \cite{AL}). The  equivalence between the $\Delta$-junction model (\ref{Y}) and the two-channel Kondo model allows one to use these results. 
 \bea
 H_{eff} = \ri\sum_p G_pS^p\epsilon_{pqt}\chi_q(0)\chi_t(0) + \frac{\ri v}{2}\int_{-\infty}^{\infty} \rd x \chi_p\p_x\chi_p , \label{Kondo}
\eea
where $G_1 = 2bJ_{23}, G_2 = 2bJ_{13}, G_3=2b J_{12}$. This description is valid at energies $<< J$. We used the equivalence between bilinears of MZEMs and components of spin S=1/2 (\ref{BS}) to replace them with the spin operator. 

  Model (\ref{Kondo}) describes the two-channel Kondo problem written in the  form  introduced in \cite{ioffe}. This equivalence is based on the fact  that the fermionic bilinears coupled to the "spin" 
\[
{\cal J}^a = \frac{\ri}{2}\epsilon_{abc}\chi_b\chi_c, 
\]
are SU$_2$(2) currents,  that satisfy the same commutation relations as the corresponding fermionic bilinears in the 2-channel Kondo model. 

 If the bare interactions are ferromagnetic ($-J_{ab}< 0$, the same sign as the bulk exchange), the Kondo exchange in (\ref{Kondo})  is antiferromagnetic and the interaction scales to the intermediate coupling critical point. Otherwise it is marginally irrelevant. A similar Kondo model (the 4-channel one) has recently been suggested in \cite{cooper} in the context of the so-called topological Kondo effect. In \cite{crampe} it was found that 4-channel Kondo model describes $\Delta$-junction of XX spin S=1/2 chains. 

As was established in \cite{ioffe}, the low energy Lagrangian describing the junction dynamics at energies less than the Kondo temperature $T_K$:
\bea
L_{eff} = \frac{\ri}{2}\epsilon\p_{\tau}\epsilon + g\epsilon\chi_1(0)\chi_2(0)\chi_3(0) + \sum_{p=1}^3L[\chi_p], \label{effTh}
\eea
where $g\sim T_K^{-1/2}$ and $L[\chi_p]$ describes three chiral Majorana modes of the bulk and $\epsilon$ is a new MZEM describing the residual degeneracy of the ground state. This MZEM is nonlocal in terms of the fields of model (\ref{Kondo}). The critical point is characterized by a single zero energy Majorana fermion coupled to the bulk by the irrelevant operator.

The scaling towards this boundary critical point takes place even if one of the couplings is zero.  Then the $\Delta$-junction becomes  a $T$-junction. Indeed, the first loop renormalization group equations for the running coupling constants are
\bea
\frac{\rd g_a}{\rd \ln\Lambda} = - g_bg_c, ~~ (b \neq c \neq a),
\eea
where $g_a(J) = 2G_a/\pi v$. Hence even if, for instance, $g_1(J) =0$, the coupling $g_1$ will be generated by two other couplings under renormalization. Thus the model flows to the 2-channel Kondo critical point  irrespectively of the ratios between the couplings $G_a$; it is well known that the multichannel Kondo model fixed  point is unaffected by the anisotropy \cite{pang}.  This is a  remarkable fact meaning that  the low energy properties of the junction are  robust with respect to anisotropy of the couplings. The Kondo scale $T_K$ corresponds to the energy  where all dimensionless coupligs $g_a$  become $\sim 1$ and is a function of the bare couplings $G_a$ and the ultraviolet cut-off $J$, for equal couplings $G_1 = G_2 = G_3$ it is exponentially small in $G$.

\section{Bethe ansatz and Thermodynamics}

 In this Appendix I consider the derivation of TBA equations for the simplest case $M=3$.  In the isotropic case 
the Bethe ansatz equations are as for $k=4$ SU(2) Kondo model with impurity spin S=1/2 \cite{wiegmann},\cite{andrei},\cite{tsvelik}:
\bea
&& [e_4(\lambda_a)]^Le_1(\lambda_a -1/g) = \prod_{b=1}^M e_2(\lambda_a -\lambda_b)\nonumber\\
&& E = \sum_{a=1}^M \frac{1}{2\ri}\ln e_4(\lambda_a), ~~ S^z = 2N - M. \label{M3}
\eea
 Here $L$ is the number of particles of the bulk, $g$ is the coupling constant, $E$ is energy. These are not equations in the scaling limit and to obtain the proper ones one has to go through some  standard motions partly described below. The Thermodynamic Bethe Ansatz  (TBA) equations in the scaling limit were  derived and the thermodynamics was studied in \cite{tsvelik}.

From (\ref{M3}) one can derive the entire thermodynamics including the results discussed in the main text. Derivation of TBA equations  follows the standard scheme. First, one establishes that generically complex solutions of (\ref{M3}) in the thermodynamic limit ($L\rightarrow \infty,  M/L =$ finite) have only fixed imaginary parts. More specifically, these solutions group into clusters with a common real part (the so-called "strings"):
    \bea
    && \lambda_{n,j;\alpha} = X_{\alpha}^{(n)} + \ri(n + 1 -2j)/2 + O(\exp(-\mbox{const} L)), \nonumber\\
&& n=1,2,...; ~ j = 1,2,...n.
    \eea
    Then  ones introduces distribution functions of rapidities of string centers $\rho_n(x)$ ( n = 1,2,...). Functions $\tilde\rho_n(x)$ describe distribution of unoccupied spaces. The discrete equations (\ref{M3}) are transformed into the integral equations relating $\tilde\rho$ and $\rho$. 
 Their ratios are parametrized by excitation energy functions $\phi_n$:
 \be
 \tilde\rho_n(x)/\rho_n(x) = \exp[-\phi_n(x)],
 \ee
\be
{\cal M}/L= \sum_{n=1}^{\infty}n\int \rd x \rho_n(x).
\ee
 The entropy of the state is given by the expression
 \be
 S = L\sum_{n=0}^{\infty}\int \rd x \left[ (\rho_n + \tilde\rho_n)\ln(\rho_n + \tilde\rho_n) - \rho_n\ln\rho_n - \tilde\rho_n\ln\tilde\rho_n\right].
 \ee
The TBA equations are a result of minimization of the generalized free energy  subject to constraints imposed by the equations for the distribution functions. The latter equations have the following form:
\bea
\tilde\rho_n + A_{nm}*\rho_m = A_{n,4}*s(x) + \frac{1}{L}a_n(x-1/g),  \label{BA3}
\eea
where 
\bea
&& A_{nm}(\omega) = \coth|\omega|/2\Big[\re^{-|n-m||\omega|/2} - \re^{-(n+m)|\omega|/2}\Big].\nonumber\\
&& a_n(\omega) = \re^{-n|\omega|/2}.
\eea
To obtain the scaling limit of (\ref{BA3}) one has to convert the kernels in such a way that they act only on the densities of holes which vanish in the ground state and after that consider the limit $x >>1, 1/g << 1$. In this limit for the bulk driving term one keeps only the main asymptotic and the impurity part $\sim 1/L$ is kept inits entirety. The result is 
\bea
\rho_n(x) + C_{nm}*\tilde\rho_m(x) = \delta_{n,4}\re^{-\pi x} + \frac{\delta_{n,1}}{L}s(x- 1/g).
\eea
where
\bea
 C_{nm} = \delta_{nm} - s(\delta_{n.m-1} + \delta_{n,m+1}),  ~~ s(\omega) = (2\cosh\omega/2)^{-1}.\nonumber
\eea
\end{appendix}

\end{document}